\documentclass[12pt]{article}
\usepackage{epsf}
\usepackage{color}
\begin{document}

\begin{flushright} 
November 2006  \\
KUNS-2040 \\
RIKEN-TH 83\\
\end{flushright} 
\vspace{0.1cm}
\begin{Large}
\vspace{1cm}
\begin{center}
{\bf Field Equations of Massless Fields 

in the New Interpretation of the Matrix Model
} \\ 
\end{center}
\end{Large}

\vspace{0.2cm}                                                                 
\begin{center}
          Ko F{\sc uruta}$^a$\footnote
           {
E-mail address : furuta@riken.jp},
          Masanori H{\sc anada}$^{b}$\footnote
           {
E-mail address : hana@gauge.scphys.kyoto-u.ac.jp}, 
          Hikaru K{\sc awai}$^{ab}$\footnote
           {
E-mail address : hkawai@gauge.scphys.kyoto-u.ac.jp} and 
          Yusuke K{\sc imura}$^{bc}$\footnote
           {
E-mail address : y.kimura@qmul.ac.uk}
%\\ and Deep Impact$^{d}$

\vspace{0.25cm}

$^a$           
{\it Theoretical Physics Laboratory,
\\
The Institute of Physical
and Chemical Research
(RIKEN),

 Wako, Saitama 351-0198, Japan}\\
$^b$           
{\it Department of Physics, Kyoto University,
Kyoto 606-8502, Japan}\\
$^c$           
{\it Department of Physics, Queen Mary, University of London, 

  Mile End Road, London E1 4NS, England}                                  
% \\
% $^d$
% {\it Ikee Kyusya, Ritto, Shiga, Japan}
\end{center}
                                                                                
\vspace{0.2cm}   

\begin{abstract}

Recently, some of the authors 
have introduced a new interpretation 
of matrix models in which 
covariant derivatives on any curved space can be expressed 
by large-$N$ matrices. 
It has been shown that the Einstein equation 
follows from the equation of motion of IIB matrix model 
in this interpretation. 
In this paper, we generalize this argument to 
covariant derivatives with  torsion.  
We find that some components of the torsion field can be 
identified with the dilaton and the $B$-field in string theory. 
However, the other components do not seem to have 
string theory counterparts. 
We also consider the matrix model with a mass term 
or a cubic term, in which the equation of motion of string theory 
is exactly satisfied. 
  
\end{abstract}

\newpage
%%%%%%%%%%%%%%%%%%%%%%%%%%%%%%%%%%%%%%%%%%%%%%%%%%%%%%%%%%
%%%%%%%%%%%%%%%%%%%%%%%%%%%%%%%%%%%%%%%%%%%%%%%%%%%%%%%%%%
%%%%%%%%%%%%%%%%%%%%%%%%%%%%%%%%%%%%%%%%%%%%%%%%%%%%%%%%%%
\section{Introduction}
%%%%%%%%%%%%%%%%%%%%%%%%%%%%%%%%%%%%%%%%%%%%%%%%%%%%%%%%%%
%%%%%%%%%%%%%%%%%%%%%%%%%%%%%%%%%%%%%%%%%%%%%%%%%%%%%%%%%%
%%%%%%%%%%%%%%%%%%%%%%%%%%%%%%%%%%%%%%%%%%%%%%%%%%%%%%%%%% 
\hspace{4.6mm}
String theory, as a candidate of unified theory of 
particle physics, is expected to provide knowledge
of physics governing high-energy regime and possibly
early time of the universe where the quantum effects
of gravity become important.
It is of crucial importance to have descriptions
of string theory which give insight beyond perturbative level.
Matrix models give efficient ways of approaching 
nonperturbative nature of string theory \cite{BFSS, IKKT}.
In matrix models, how to describe the spacetime geometry 
has been one of key questions.

In the original proposal of matrix models, matrices have 
been considered as spacetime coordinates.
In this description, the gravitational interaction 
appears as a loop effect, but the origin of the general covariance 
is unclear.   
In the recent paper \cite{HKK}, another interpretation 
of matrix models was proposed by some of the authors. 
It has been shown that 
a covariant derivative on any $d$ dimensional spacetime $M$  
can be expressed by a set of $d$ operators 
that acts on the space of functions on the principal $Spin(d)$ 
bundle over $M$. 
The Einstein equation follows from the equation of motion 
of the matrix model if we identify matrices 
with the covariant derivative in this way. 
Symmetries under local Lorentz transformation
and diffeomorphism are manifestly realized as 
a part of the $U(N)$ symmetry. 
On the other hand, a general matrix should be regarded 
as a power series in the covariant derivative and Lorentz generator 
that contains infinitely many fields as coefficients. 
To understand their dynamics, it is important to 
investigate the degrees of freedom which appear 
around
a classical solution. 
In this paper, we study the fields 
that appear as the coefficients of terms linear in 
the covariant derivative and Lorentz generator, 
that is, the vielbein and the torsion\footnote{
Higher spin fields appear 
from coefficients of higher powers of covariant derivatives, and 
their gauge symmetries are included in the $U(N)$ symmetry \cite{HKK}. 
Free equations of motion are also correctly reproduced 
from that of the matrix model \cite{Saitou}. 
}. 
We discuss a possible interpretation of these fields in 
the framework of string theory. 

The paper is organized as follows.
In the next section, we review the new interpretation 
of matrix models \cite{HKK}. In particular, 
we explain how the covariant derivative on a curved space 
can be expressed in terms of matrices.   
In section \ref{sec:Torsion_as_massless_fields}, 
we analyze the dynamics of the vielbein and the torsion. 
They form a closed subset of equations 
of motion and are expected to have consistent solutions
without introducing higher spin fields.
By identifying the trace and the antisymmetric parts of the torsion 
to be the field strengths of the dilaton and the antisymmetric 
$B$-field, respectively, 
we find solutions which resemble those in string theory. 
However, it turns out that ghost like degrees of freedom 
can also propagate in general. 
We might need to 
impose a constraint on the expansion of matrices. 
In section \ref{sec:modified_action}, we study modified models 
which have mass and cubic terms. 
In these cases, there are classical solutions with constant torsion. 
If we interpret the antisymmetric part of the torsion to be 
the field strength of the $B$-field, then these solutions satisfy  
the Einstein equation with a constant background of field strength 
of $B$. 
This supports the idea of embedding the $B$-field into the 
torsion that is introduced 
in section \ref{sec:Torsion_as_massless_fields}.
Section \ref{sec:Discussions} is devoted to conclusions 
and discussions on future directions.
Some of the details are contained in the appendices.
%%%%%%%%%%%%%%%%%%%%%%%%%%%%%%%%%%%%%%%%%%%%%%%%%%%%%%%%%%
%%%%%%%%%%%%%%%%%%%%%%%%%%%%%%%%%%%%%%%%%%%%%%%%%%%%%%%%%%
%%%%%%%%%%%%%%%%%%%%%%%%%%%%%%%%%%%%%%%%%%%%%%%%%%%%%%%%%%
\section{New Interpretation of Matrix Model}
\label{sec:Matrix representation of covariant derivative}
%%%%%%%%%%%%%%%%%%%%%%%%%%%%%%%%%%%%%%%%%%%%%%%%%%%%%%%%%%
%%%%%%%%%%%%%%%%%%%%%%%%%%%%%%%%%%%%%%%%%%%%%%%%%%%%%%%%%%
%%%%%%%%%%%%%%%%%%%%%%%%%%%%%%%%%%%%%%%%%%%%%%%%%%%%%%%%%%
\hspace{4.6mm}
In this section, we briefly summarize the results obtained in \cite{HKK}.
We consider the large-$N$ reduced model of $d$-dimensional 
$U(N)$ Yang-Mills theory,
\begin{equation}\label{reducedmodel}
  S
  =
  -\frac{1}{4g^2}Tr
  \biggl(\left[A_a,A_b\right]\left[A^a,A^b\right]
  \biggr), 
\end{equation}
where $A_a$ are $N\times N$ 
Hermitian matrices, and $a$ and $b$ run from $1$ to $d$. 
(Supersymmetric version of this model
with $d=10$ is known as IIB matrix model \cite{IKKT}.)
This model has the $U(N)$ gauge symmetry
\begin{eqnarray}
  \delta A_a=i[\Lambda,A_a], 
\end{eqnarray}
where $\Lambda$ is a Hermitian matrix. 

In \cite{HKK}, a new interpretation is proposed  
in which the matrices $A_a$ represent differential operators 
on a Riemannian manifold. 
In this interpretation, 
the symmetries of general relativity are contained naturally 
in the $U(N)$ symmetry, and 
the Einstein equation follows from the equation of motion 
of the matrix model. 

To begin with, let us see 
how the covariant derivatives are described in terms of matrices. 
An important aspect of the covariant derivative is 
that it maps a rank-$n$ tensor to a rank-$(n+1)$ tensor. 
For example, $\nabla_a V_b$ is not the $a$-th component\footnote{
$a,b,\cdots$ represent the Lorentz indices. 
We use $\mu,\nu,\cdots$ to express the Einstein indices.  
} of $\nabla$ 
acting on the $b$-th component of $V$,  
but is the $(a,b)$-component of a rank-two tensor $\nabla V$; 
indeed, in the formula
\begin{eqnarray}
  \nabla_a V_b
  =
  e_a{}^\mu
  \left(
    \partial_\mu V_b
    +
    \omega_{\mu b}{}^c V_c
  \right),
\end{eqnarray}
all of the components $V_c$ contribute to 
the covariant derivative.
Here, $e_a{}^\mu$ and $\omega_\mu{}^{bc}$ are 
the vielbein and the spin connection, respectively. 

Therefore, in order to express the covariant derivative 
in terms of matrices, 
the space on which matrices act 
must contain the tensors of all ranks.
%In \cite{HKK}, 
It was shown that the covariant derivative
can be interpreted as a set of matrices 
(or equivalently, endomorphisms) 
acting on functions on the principal $Spin(d)$ bundle over 
the Riemannian manifold $M_d$. 
The space of functions on the group manifold $Spin(d)$ 
forms the regular representation which contains  
tensors of all ranks. Then, 
the covariant derivative $\nabla_a$ is mapped to
a set of matrices $\nabla_{(a)}$ by 
\begin{eqnarray}
  \nabla_{(a)}
  &=& 
  R_{(a)}{}^b(g^{-1})\nabla_b
  \nonumber\\
  &=&
  R_{(a)}{}^b(g^{-1}) e_b{}^\mu(x)
  \left(
    \partial_\mu+\omega_\mu{}^{cd}(x) {\cal O}_{cd} 
    % -ia_\mu(x)
  \right),
\end{eqnarray}
where $g$ is the coordinate of the fiber $Spin(d)$, 
$R_{(a)}{}^b(g^{-1})$ is the vector representation 
of $Spin(d)$  
and  ${\cal O}_{ab}$ is the Lorentz generator of the $Spin(d)$ group. 
The operator $\nabla_{(a)}$ acts as endomorphism 
on the space of functions on the principal $Spin(d)$ bundle.
Products of $\nabla_{(a)}$ are related to those of $\nabla_a$ as follows:
\begin{eqnarray}
  \nabla_{(a)}\nabla_{(b)}\nabla_{(c)}\cdots
  &=&
  R_{(a)}{}^{a'}(g^{-1})R_{(b)}{}^{b'}(g^{-1})R_{(c)}{}^{c'}(g^{-1})
  \nabla_{a'}\nabla_{b'}\nabla_{c'}\cdots.
  \label{rule_for_omitting_parenthesis}
\end{eqnarray}
On the left hand side, each of the indices $(a),(b),(c)$
is nothing more than a label and there is no mixing of them 
in the product. Thus the product can be represented as 
the product of matrices $\nabla_{(a)},\nabla_{(b)},\nabla_{(c)}, \cdots$.
On the right hand side, product of 
$\nabla_{a^\prime},\nabla_{b^\prime},\nabla_{c^\prime},\cdots$
mixes the indices
$b^\prime, c^\prime,\cdots$.

We regard that the matrices $A_a$ act on the space of 
functions on the principal 
$Spin(d)$ bundle over $M_d$. 
Then, $A_a$ can be expanded 
as series in powers of $\nabla_{(a)}$ and ${\cal O}_{ab}$,~
\footnote{
The coefficients $a_{(a)}(x,g),b_{(a)}^{(b)}(x,g),\cdots$ 
are real-valued. 
We introduce the anticommutator $\{\ ,\ \}$ 
in order to make $A_a$ Hermitian. 
Hereafter, we omit the anticommutator for simplicity. 
}
\begin{eqnarray}
  A_a
  &=&
  a_{(a)}(x,g)
  +\frac{i}{2}
  \{
  \delta_{(a)}{}^{(b)}
  +
  b_{(a)}{}^{(b)}(x,g),\nabla_{(b)}\}
  +\frac{i}{2}\{c_{(a)}{}^{bc}(x,g),{\cal O}_{bc}\}
  \nonumber\\
  & &
  +\frac{i^2}{2}\{d_{(a)}{}^{((b)(c))}(x,g),\nabla_{(b)}\nabla_{(d)}\}
  +\frac{i^2}{4}\{\{e_{(a)}{}^{bc(d)}(x,g),{\cal O}_{bc},\},\nabla_{(d)}\}
  +\cdots.       
  \label{expansion}
\end{eqnarray}
If we interpret the coefficients appearing in this expansion 
to be local fields, 
then the diffeomorphism, the local Lorentz transformation 
and higher-spin gauge transformations are contained naturally 
in the $U(N)$ symmetry of the matrix model \cite{HKK,Saitou}. 
For example, the diffeomorphism is generated by 
$\Lambda=i\lambda^{(a)}(x)\nabla_{(a)}$.
(We will show the explicit expression for the diffeomorphism and 
the local Lorentz transformation 
in section \ref{sec:Torsion_as_massless_fields}. )  

The equation of motion for the action (\ref{reducedmodel})
becomes
\begin{eqnarray}
  [A^a,[A_a,A_b]]=0.
  \label{eom_matrix}
\end{eqnarray}
As an ansatz, it is natural to require $A_a$ to be the differential 
operators of first-order, because their commutator is again 
of first-order and hence they form a closed algebra. 
Here, we consider the simplest possibility among them,  
\begin{eqnarray}
  A_a=i\nabla_{(a)}. 
  \label{ansatz}
\end{eqnarray}
(We consider a general ansatz in the next section.)  
Then, (\ref{eom_matrix}) becomes 
\begin{eqnarray}\label{eom gravity}
  [\nabla^a,[\nabla_a,\nabla_b]]
  =(\nabla^a R_{ab}{}^{cd}){\cal O}_{cd}
  -R_b{}^c\nabla_c
  =0, 
\end{eqnarray}
where we have converted the indices $(a)$ and $(b)$ 
to the indices $a$ and $b$ using the equation 
(\ref{rule_for_omitting_parenthesis}). 
Because 
$\nabla^a R_{ab}{}^{cd}=0$ 
follows from $R_{ab}=0$ and the Bianchi identity 
$\nabla^{[a}R_{ab}{}^{cd]}=0$, 
the equation (\ref{eom gravity}) reduces to 
$R_{ab}=0$ and thus provides the Einstein equation. 
Furthermore, if we regard the fields appearing in  
(\ref{expansion}) to be fluctuations about 
the classical solution of the form (\ref{ansatz}), 
they can be interpreted as local fields propagating on 
the curved space $M_d$ which is represented by 
the covariant derivative $\nabla_{(a)}$; for example, 
$a_a(x)$ satisfies the Maxwell equation \cite{HKK} and  
$d_a{}^{bc}$ satisfies the free equations of motion 
for higher-spin fields \cite{Saitou}. 
Therefore, we can interpret that the classical solution 
derived above describes the Ricci-flat background spacetime. 
This is a new mechanism of the emergence of the spacetime 
in the matrix model. 

% Here we comment on the $g$ dependence of 
% fields $a_a(x,g)$, $b_a{}^b(x,g), \cdots$. 
% Because there seem to appear too many fields due to this 
% $g$ dependence, it would be better to restrict our matrix space 
% by imposing a constraint. 
% We introduce the right action of $Spin(d)$ as 
% \begin{equation}
%   a_{(a)}(x,g)\rightarrow a_{(a)}(x,gh^{-1}). 
% \end{equation}
% We require that $a_{(a)}(x,g)$ transforms as a vector under 
% this action, 
% \begin{equation}
%   a_{(a)}(x,gh^{-1})=R_{(a)}{}^{(b)}(h)a_{(b)}(x,g) .
%   \label{transformationunderrightaction}
% \end{equation}
% Because $a_{(a)}(x,gh^{-1})$ can be rewritten 
% in the following way, 
% \begin{eqnarray}
% a_{(a)}(x,gh^{-1})=R_{(a)}{}^{b}(hg^{-1})a_{b}(x,gh^{-1})
% =R_{(a)}{}^{b}(h) R_{b}{}^{c}(g^{-1})a_{c}(x,gh^{-1}),  
% \end{eqnarray}
% this requirement provides a constraint 
% $a_{b}(x,g)=a_{b}(x,gh^{-1})$, which means 
% the $g$-independece of $a_{b}(x,g)$. In this way, 
% we can restrict the matrix space 
% to a smaller one.\footnote{
% A detailed discussion is given in section 3.2 of \cite{HKK}. }
% An issue related to a constraint on the matrix space is also 
% discussed in section \ref{sec:Discussions}.
%%%%%%%%%%%%%%%%%%%%%%%%%%%%%%%%%%%%%%%%%%%%%%%%%%%%%%%%%%
%%%%%%%%%%%%%%%%%%%%%%%%%%%%%%%%%%%%%%%%%%%%%%%%%%%%%%%%%%
%%%%%%%%%%%%%%%%%%%%%%%%%%%%%%%%%%%%%%%%%%%%%%%%%%%%%%%%%%
\section{Torsion as Massless Fields}\label{sec:Torsion_as_massless_fields}
%%%%%%%%%%%%%%%%%%%%%%%%%%%%%%%%%%%%%%%%%%%%%%%%%%%%%%%%%%
%%%%%%%%%%%%%%%%%%%%%%%%%%%%%%%%%%%%%%%%%%%%%%%%%%%%%%%%%%
%%%%%%%%%%%%%%%%%%%%%%%%%%%%%%%%%%%%%%%%%%%%%%%%%%%%%%%%%%
\hspace{4.6mm}
In this section, 
we consider the general form of 
the first-order differential operator, which is given by 
\begin{eqnarray}\label{ansatz1}
  A_a
  =
  iD_{(a)}
  \equiv
  iR_{(a)}{}^b(g^{-1})D_b
  \equiv
  iR_{(a)}{}^b(g^{-1})\left(
    e_b{}^\mu(x)\nabla_\mu 
    +
    S_{b}{}^{cd}(x){\cal O}_{cd}
  \right), 
\end{eqnarray}
where $\nabla_\mu$ and $S_a{}^{bc}(x)$ are 
the torsionless covariant derivative and the contorsion, 
respectively.\footnote{
  Although we can introduce coefficients 
  of the covariant derivative as 
  $A_a=i\left(a_{(a)}{}^{(b)}\nabla_{(b)}
    +S_{(a)}{}^{bc}{\cal O}_{bc}\right)$, 
  such coefficients can be absorbed by the following field
  redefinition:
  $e^\prime_{(a)}{}^\mu=
  a_{(a)}{}^{(b)}e_{(b)}{}^\mu, 
  S^\prime_{(a)}{}^{bc}{\cal O}_{bc}
  =
  S_{abc}{\cal O}_{bc}+
  \left(
    a_{(a)}{}^{(b)}\nabla_{(b)}-\nabla^\prime_{(a)}
  \right). $
Here $\nabla^\prime_a$ is the torsionless covariant derivative 
which is constructed from $e^\prime_a{}^\mu$. 
}
The differential operator $D_a$ 
can be regarded as a covariant derivative with 
torsion.\footnote{
The torsion gravity has been studied extensively 
as a generalization of general relativity, and 
it is interesting to clarify the relation of it to 
our approach. For reviews of the torsion gravity, see 
\cite{HHKN,Hammond}. 
}
As discussed in the previous section, 
because differential operators of first-order form a closed algebra,  
it is expected that the equation of motion has consistent 
solutions by setting all the fields in (\ref{expansion})
to be zero except for the fields
appearing in the coefficients of (\ref{ansatz1}).
Note that it is possible to consider $g$-dependent
fields $e_b{}^{\mu}(x,g)$ and $S_b{}^{cd}(x,g)$ in general.
In this paper, we restrict our attention to the simplest
case where $e_b{}^\mu$ and $S_b{}^{cd}$ has
no $g$-dependence. 
For details concerning this point, see \cite{HKK}.

Before analyzing the equation of motion, 
it is helpful to study the gauge symmetries. 
The action (\ref{reducedmodel}) is invariant under 
the $U(N)$ transformation $\delta A_a=i[\Lambda,A_a]$, 
where $\Lambda$ is an $N\times N$ Hermitian matrix. 
If we restrict $\Lambda$ to be the first-order differential operator, 
\begin{eqnarray}
  \Lambda
  =
  i\lambda^\mu(x)\nabla_\mu
  +
  i\lambda^{ab}(x){\cal O}_{ab}, 
\end{eqnarray}
then the local fields appearing in (\ref{ansatz1}) transform as 
\begin{eqnarray}
  & &
  \delta e_a{}^\mu(x)
  =
  -\lambda_a{}^b(x)e_b{}^\mu(x)
  +
  \nabla_a\lambda^\mu(x), 
  \\
  & &
  \delta \omega_\mu{}^{bc}(x)
  =
  \nabla_\mu\lambda^{bc}(x)
  +
  \lambda^\nu(x)R_{\mu\nu}{}^{bc}(x), 
  \\
  & &
  \delta S_a{}^{bc}
  =
  -
  \lambda^\mu(x)\nabla_\mu S_a{}^{bc}(x)
  -
  \lambda_a{}^d S_d{}^{bc}(x)
  -
  \lambda^b{}_d S_a{}^{dc}(x)
  -
  \lambda^c{}_d S_a{}^{bd}(x). 
  \label{Sisnotgaugefield}
\end{eqnarray}
Therefore, $i\lambda^\mu(x)\nabla_\mu$ and 
$i\lambda^{ab}(x){\cal O}_{ab}$ generate the diffeomorphism and 
the local Lorentz transformation, respectively, 
and we can use them to remove unphysical modes 
of $e_a{}^\mu(x)$. 
Then, no component of the contorsion $S_{a}{}^{bc}(x)$ 
can be gauged away. 

Now, let us derive the equation of motion. 
The commutator of $D_a$ gives
\begin{eqnarray}
  [D_a,D_b]={\cal R}_{ab}{}^{cd}{\cal O}_{cd}+T_{ab}{}^c D_c, 
  \label{definition of cal R and T}
\end{eqnarray}
where $T_{abc}=S_{abc}-S_{bac}$ is the torsion 
and ${\cal R}_{ab}{}^{cd}$ is the Riemann tensor 
with the contribution from the torsion, 
which is related to the ordinary Riemann tensor $R_{ab}{}^{cd}$ 
by 
\begin{eqnarray}
  {\cal R}_{ab}{}^{cd}
  =
  R_{ab}{}^{cd}
  +
  \nabla_a S_b{}^{cd}
  -
  \nabla_b S_a{}^{cd}
  -
  S_{ae}{}^{c}S_b{}^{ed}
  +
  S_{be}{}^{c}S_a{}^{ed}. 
  \label{Riemann_tensor_with_torsion}
\end{eqnarray}
The equation of motion is given by
\begin{eqnarray}\label{eom torsion}
  [D^a,[D_a,D_b]]
  &=&
  \left(
    -{\cal R}_{b}{}^c
    +
    D^a T_{ab}{}^c
    +
    T_{ab}{}^d T^a{}_{d}{}^c
  \right)D_c
  \nonumber\\
  & &
  +
  \left(
    D^a{\cal R}_{ab}{}^{cd}
    +
    T^a{}_b{}^e {\cal R}_{ae}{}^{cd}
  \right){\cal O}_{cd}
  \nonumber\\
  &=&
  0, 
\end{eqnarray}
where ${\cal R}_{bc}={\cal R}_{ab}{}^a{}_c$. 
Therefore, we obtain the following two equations: 
\begin{eqnarray}
  & &
  -{\cal R}_{bc}
  +
  D^a T_{abc}
  +
  T_{ab}{}^d T^a{}_{dc}
  =
  0,
  \label{nabla-eq R T} \\
  & &
  D^a{\cal R}_{ab}{}^{cd}
  +
  T^a{}_b{}^e {\cal R}_{ae}{}^{cd}
  =
  0.
  \label{O-eq R T}
\end{eqnarray}
Note that (\ref{nabla-eq R T}) has not only the symmetric part, 
which can be regarded as the Einstein equation, 
but also the antisymmetric part. 
These equations are solved in the following subsections. 
We first analyze them at the linearized level   
in the next subsection. 
We find solutions in which torsion fields 
can be interpreted as 
the dilaton and the antisymmetric $B$-field in string theory. 
In subsection \ref{The existence of solutions} we show that 
the equations have solutions
to all orders in a weak field expansion.   
%%%%%%%%%%%%%%%%%%%%%%%%%%%%%%%%%%%%%%%%%%%%%%%%%%%%%%%%%%
%%%%%%%%%%%%%%%%%%%%%%%%%%%%%%%%%%%%%%%%%%%%%%%%%%%%%%%%%%
%%%%%%%%%%%%%%%%%%%%%%%%%%%%%%%%%%%%%%%%%%%%%%%%%%%%%%%%%%
\subsection{Linear Approximation}\label{subsec:linear}
%%%%%%%%%%%%%%%%%%%%%%%%%%%%%%%%%%%%%%%%%%%%%%%%%%%%%%%%%%
%%%%%%%%%%%%%%%%%%%%%%%%%%%%%%%%%%%%%%%%%%%%%%%%%%%%%%%%%%
%%%%%%%%%%%%%%%%%%%%%%%%%%%%%%%%%%%%%%%%%%%%%%%%%%%%%%%%%%
\hspace{4.6mm}
In this subsection, we solve the equations of motion 
(\ref{nabla-eq R T}) and (\ref{O-eq R T}) 
at the linearized level. 
The contorsion 
$S_a{}^{bc}$ can be decomposed to irreducible tensors 
as 
\begin{eqnarray}
  S_{abc}
  =
  V_b\delta_{ac}
  -
  V_c\delta_{ab}
  +
  L_{abc}
  +
  H_{abc},   
\end{eqnarray}
where $H_{abc}$ is antisymmetric and $L_{abc}$ satisfies 
the relations $L_{[abc]}=0$ and $ L^a{}_{ac}=0$. 
Using these irreducible tensors, 
we can linearize (\ref{nabla-eq R T}) and (\ref{O-eq R T})
about the flat space, and we obtain  
\begin{eqnarray}
  & &
  R_{ab}
  =
  (3-d)\partial_{(a}V_{b)}
  -
  2(\partial\cdot V)\delta_{ab}
  +
  2\partial^c L_{(ab)c}, 
  \label{EOM_linear_Einstein}\\
  %%%%%%%%%%%%%%% 
  & &
  (1-d)\partial_{[a}V_{b]}
  +
  3\partial^c H_{abc}
  =
  0, 
  \label{EOM_linear_antisymmetric}\\
  %%%%%%%%%%%%%%%%
  & &
  \partial_a
  \left(
    R_{ab}{}^{cd}
    +
    \partial_a S_{b}{}^{cd}
    -
    \partial_b S_{a}{}^{cd}
  \right)
  =
  0. 
  \label{EOM_linear_O_ij}
\end{eqnarray}
Note that here $\partial_a$ represents
$\delta_a{}^\mu\partial_\mu$. The expressions  
(\ref{EOM_linear_Einstein}) and (\ref{EOM_linear_antisymmetric}) are 
the symmetric and antisymmetric part of (\ref{nabla-eq R T}),
respectively. 
%In general, all components of the torsion field can propagate. 
As we have discussed below (\ref{Sisnotgaugefield}), 
$V_a$, $H_{abc}$ and $L_{abc}$ are not gauge fields. 
Therefore, if we regard them as fundamental fields, 
there appear negative-norm modes.\footnote
{By negative modes we mean fields with
odd number of time like index $0$, 
although we have not started from any field theory action. 
It is hard to expect such degrees of freedom to 
posses positive norm when the equation
of motion could be derived from a field theory action. } 
Then, it is natural to regard them as field strengths. 
As we will see below, this assumption 
leads to equations of motion 
of massless fields in string theory at least in the linearized level. 
For this reason, we consider 
the following form of solutions:
% \footnote{
%   Instead of (\ref{ansatz_at_linearized_level_L}) 
%   we can also impose $L_{abc}=\partial_a\partial_{[b}C_{c]}$, where 
%   $C_a$ satisfies the standard equation of motion 
%   $\partial^a \partial_{[a}C_{b]}=0$,  
%   without altering (\ref{eom_linear_Einstein}), 
%   (\ref{eom_linear_phi}) and (\ref{eom_linear_B}).   
% }
\begin{eqnarray}
  & &
  V_a
  =
  \partial_a\phi, 
  \label{ansatz_at_linearized_level_V}
  \\
  & &
  H_{abc}
  =
  \partial_{[a}B_{bc]}, 
  \label{ansatz_at_linearized_level_H}
  \\
  & &
  L_{abc}
  =
  0. 
  \label{ansatz_at_linearized_level_L}
\end{eqnarray}
Then, the equations of motion reduce to 
\begin{eqnarray}
  & &
  R_{ab}
  =
  (3-d)\partial_a\partial_b\phi, 
  \label{eom_linear_Einstein}
  \\
  & &
  \partial^2\phi
  =
  0, 
  \label{eom_linear_phi}
  \\
  & &
  \partial^a H_{abc}
  =
  0. 
  \label{eom_linear_B}
\end{eqnarray}
Here, the equation (\ref{eom_linear_B})
is obtained directly from 
(\ref{EOM_linear_antisymmetric}).  
The equation (\ref{eom_linear_phi}) 
is obtained by substituting 
(\ref{EOM_linear_Einstein})  to the Bianchi identity 
$\nabla_b R=2\nabla^a R_{ab}$. 
Using the equation (\ref{eom_linear_phi}), 
we obtain the equation (\ref{eom_linear_Einstein}). 
Using the Bianchi identities 
$\nabla^{[a}R_{ab}{}^{cd]}=0$ and 
$\partial_{[a} H_{bcd]}=0$, we can see that 
(\ref{EOM_linear_O_ij}) gives no other equation.    

These equations are the same as 
the linearized equations of motion 
for the system of the dilaton and 
the antisymmetric $B$-field coupled to gravity, 
which arise naturally from the string theory.\footnote{
An earlier attempt to embed the $B$-field in torsion 
can be found in \cite{SS}. 
}  
Note that the conditions 
(\ref{ansatz_at_linearized_level_V}), 
(\ref{ansatz_at_linearized_level_H}) and 
(\ref{ansatz_at_linearized_level_L}) 
are modified if we include the nonlinear terms. 
We discuss this point in the next subsection. 

Although we have considered solutions which satisfy 
the conditions 
(\ref{ansatz_at_linearized_level_V}), 
(\ref{ansatz_at_linearized_level_H}) and 
(\ref{ansatz_at_linearized_level_L}), 
as we will see in the next subsection,
%f1
and in appendix \ref{appendix:DOF}, 
there are many solutions which do not satisfy these 
conditions. 
Such solutions are undesirable, because 
negative norm states can not be removed by the gauge symmetry.  
We discuss this point further in section \ref{sec:Discussions}. 
%%%%%%%%%%%%%%%%%%%%%%%%%%%%%%%%%%%%%%%%%%%%%%%%%%%%%%%%%%
%%%%%%%%%%%%%%%%%%%%%%%%%%%%%%%%%%%%%%%%%%%%%%%%%%%%%%%%%%
%%%%%%%%%%%%%%%%%%%%%%%%%%%%%%%%%%%%%%%%%%%%%%%%%%%%%%%%%%
\subsection{Weak Field Expansion}\label{The existence of solutions}
%%%%%%%%%%%%%%%%%%%%%%%%%%%%%%%%%%%%%%%%%%%%%%%%%%%%%%%%%%
%%%%%%%%%%%%%%%%%%%%%%%%%%%%%%%%%%%%%%%%%%%%%%%%%%%%%%%%%%
%%%%%%%%%%%%%%%%%%%%%%%%%%%%%%%%%%%%%%%%%%%%%%%%%%%%%%%%%%
\hspace{4.6mm}
In this subsection, we introduce a systematic weak field expansion. 
We find that for any solution of the linearized equations 
we can determine 
higher order corrections order by order. 
In particular, the solution we discussed in the previous subsection, 
which looks like the system of the dilaton and 
the $B$-field coupled to gravity, 
can be lifted to all orders.  

In order to perform the weak field expansion,  
it is convenient to redefine the fields as 
\begin{eqnarray}
  D_a
  =
  e_a{}^\mu\partial_\mu
  +
  S^\prime_a{}^{bc}{\cal O}_{bc}, 
\end{eqnarray}
where the quantity $S^\prime_a{}^{cd}$ is a sum of the contorsion 
and the spin connection, 
\begin{eqnarray}
  S^\prime_a{}^{bc}
  =
  S_a{}^{bc}+e_a{}^\mu\omega_\mu{}^{bc}. 
\end{eqnarray}
By expanding $e_a{}^\mu$ about the flat space as 
\begin{eqnarray}
  e_a{}^\mu
  =
  \delta_a{}^\mu+h_a{}^\mu, 
\end{eqnarray}
the equations of motion can be written as
\begin{eqnarray}
  \left[D^a,[D_a,D_b]\right]
  =
  \left(
    \partial^a f_{ab}{}^{\mu}
    -
    j_{b}{}^{\mu}
  \right)
  \partial_\mu
  +
  \left(
    \partial^a F_{ab}{}^{cd}
    -
    J_{b}{}^{cd}
  \right){\cal O}_{cd}
  =
  0,    
  \label{proof_of_existence_of_solution_0}
\end{eqnarray}
where 
\begin{eqnarray}
  f_{ab}{}^{\mu}
  =
  \partial_{a}h_{b}{}^\mu
  -
  \partial_{b}h_{a}{}^\mu, 
  \qquad
  F_{ab}{}^{cd}
  =
  \partial_{a}S^\prime_{b}{}^{cd}
  -
  \partial_{b}S^\prime_{a}{}^{cd},  
  \label{field strength}
\end{eqnarray}
and 
$j_a{}^\mu$ and $J_a{}^{cd}$ are the polynomials of 
$h_a{}^\mu$ and $S^\prime_a{}^{cd}$.
Note that all the indices 
$a,b,\mu,\nu,\cdots$ 
are those in the flat space. 
In particular, the derivative $\partial_a$ 
does not stand for $e_a{}^\mu\partial_\mu$ but  
$\delta_a{}^\mu\partial_\mu$. 
The quantity $J_a{}^{cd}$ has only the terms of degree two and three 
in $h_a{}^\mu$ and $S_a^{\prime}{}^{cd}$, 
while the quantity $j_b{}^\mu$ has also the linear term;  
\begin{eqnarray}
  j_b{}^\mu
  =  
  -\frac{1}{2}\partial^a S^{\prime}_{ab}{}^\mu
  +
  \partial^a S^{\prime}_{ba}{}^\mu
  -
  \frac{1}{2}\partial_b S^{\prime}_{a}{}^{a\mu}
  +
  \mbox{terms of degree 2 and 3}. 
  \label{eq:small_j}
\end{eqnarray}
Because $\partial_\mu$ and ${\cal O}_{cd}$ are independent operators, 
the equation (\ref{proof_of_existence_of_solution_0}) is equivalent to 
the following equations: 
\begin{eqnarray}
  & &
  \left(
    \left.
      \left[D^a,[D_a,D_b]\right]
    \right|_{\partial}
  \right)^\mu
  =
  \partial^a f_{ab}{}^{\mu}
  -
  j_{b}{}^{\mu}
  =
  0, 
  \label{proof_of_existence_of_solution_1}
  \\
  & &
  \left(
    \left.
      \left[D^a,[D_a,D_b]\right]
    \right|_{{\cal O}}
  \right)^{cd}
  =
  \partial^a F_{ab}{}^{cd}
  -
  J_{b}{}^{cd}
  =
  0.     
  \label{proof_of_existence_of_solution_2}
\end{eqnarray}
These expressions look similar to Yang-Mills theory 
except that there are extra indices $\mu, c$ and $d$. 

Let us solve equations    
(\ref{proof_of_existence_of_solution_1}) and 
(\ref{proof_of_existence_of_solution_2}) order by order 
in a weak field expansion. 
We expand $h_a{}^\mu$ and $S^\prime_a{}^{bc}$ as 
\begin{eqnarray}
  h_a{}^\mu
  =
  h^{(1)}_a{}^\mu
  +
  h^{(2)}_a{}^\mu
  +
  \cdots, 
  \qquad
  S^\prime_a{}^{bc}
  =
  S^{\prime(1)}_a{}^{bc}
  +
  S^{\prime(2)}_a{}^{bc}
  +
  \cdots, 
\end{eqnarray}
where $h^{(i)}_a{}^\mu$ and $S^{\prime(i)}_a{}^{bc}$ represent 
quantities of $i$-th order. 
The equations of motion at the $i$-th order are 
\begin{eqnarray}
  \left(
    \left.
      \left[D^a,[D_a,D_b]\right]^{(i)}
    \right|_{\partial}
  \right)^\mu
  =
  \partial^a f_{ab}^{(i)\mu}
  -
  j_{b}^{(i)\mu}
  \left(
    h^{(1)},\cdots,h^{(i-1)},
    S^{\prime(1)},\cdots S^{\prime(i-1)},S^{\prime(i)}
  \right)
  =
  0, 
  \label{EOM:partial}
  \\
  \left(
    \left.
      \left[D^a,[D_a,D_b]\right]^{(i)}
    \right|_{{\cal O}}
  \right)^{cd}
  =
  \partial^a F_{ab}^{(i)cd}
  -
  J_{b}^{(i)cd}
  \left(
    h^{(1)},\cdots,h^{(i-1)},
    S^{\prime(1)},\cdots, S^{\prime(i-1)}
  \right)
  =
  0.   
  \label{EOM:O}
\end{eqnarray}
Note that $j_a^{(i)\mu}$ contains the contribution 
from $S_a^{\prime(i)cd}$, while $J_a{}^{(i)cd}$ does not. 

At the lowest order (\ref{EOM:O}) becomes  
\begin{eqnarray}
  \partial^a F_{ab}^{(1)cd}
  =
  0, 
  \label{existence_of_solution_1st_order_1}
\end{eqnarray}
because $J_a^{(1)cd}=0$. 
This is nothing but the Maxwell equation with the extra indices 
$c$ and $d$, and can be solved for $S_a^{\prime(1)cd}$. 
By substituting this solution $S_a^{\prime(1)cd}$ 
to (\ref{EOM:partial}), we obtain at the lowest order 
\begin{eqnarray}
  \partial^a f_{ab}^{(1)\mu}
  =
  j_b^{(1)\mu}\left(S^{\prime (1)}\right).   
  \label{eq:EOM_partial_lowest_order}
\end{eqnarray} 
We can show that 
the divergence of $j_b^{(1)\mu}$ vanishes automatically. 
Indeed, 
using (\ref{eq:small_j}) and 
(\ref{existence_of_solution_1st_order_1}), we have 
\begin{eqnarray}
  \partial^b j_b^{(1)\mu}\left(S^{\prime (1)}\right)
  =
  -
  \frac{1}{2}\left(
    \partial^2 S^{\prime(1)}_a{}^{a\mu}
    -
    \partial^a \partial^b S^{\prime(1)}_{ab}{}^\mu
  \right)
  =
  -\frac{1}{2}\partial^a F_{ab}^{(1)b\mu}
  =
  0. 
\end{eqnarray}
Therefore, (\ref{eq:EOM_partial_lowest_order}) 
has the same structure as the Maxwell equation 
with a conserved source except for an extra index $\mu$, 
and has a solution for $h_a^{(1)\mu}$. 

In this way, we have solved the equation of motion 
at the lowest order. 
We next show that the higher order terms can be determined 
order by order. 
Suppose we have solved the equation of motion to 
the $(n-1)$-th order: 
\begin{eqnarray}
  \left[D^a,[D_a,D_b]\right]^{(i)}=0
  \qquad(i=1,\cdots,n-1). 
  \label{EOM:lower order}
\end{eqnarray}
Under this assumption, we show that 
the equation of motion at the $n$-th order, 
\begin{eqnarray}
  \left[D^a,[D_a,D_b]\right]^{(n)}=0, 
\end{eqnarray}
or equivalently 
\begin{eqnarray}
  \left(
    \left.
      \left[D^a,[D_a,D_b]\right]^{(n)}
    \right|_{\partial}
  \right)^\mu
  =
  \partial^a f_{ab}^{(n)\mu}
  -
  j_{b}^{(n)\mu}
  \left(
    h^{(1)},\cdots,h^{(n-1)},
    S^{\prime(1)},\cdots S^{\prime(n-1)},S^{\prime(n)}
  \right)
  =
  0, 
  \label{EOM:partial_n-th_order}
  \\
  \left(
    \left.
      \left[D^a,[D_a,D_b]\right]^{(n)}
    \right|_{{\cal O}}
  \right)^{cd}
  =
  \partial^a F_{ab}^{(n)cd}
  -
  J_{b}^{(n)cd}
  \left(
    h^{(1)},\cdots,h^{(n-1)},
    S^{\prime(1)},\cdots, S^{\prime(n-1)}
  \right)
  =
  0,    
  \label{EOM:O_n-th_order}  
\end{eqnarray}
also has a solution for $h_a^{(n)\mu}$ and $S_a^{\prime(n)cd}$. 
What we need is to show 
\begin{eqnarray}
  & &
  \partial^b J_b^{(n)cd}=0,
  \label{conservation_large_J}
  \\
  & &
  \partial^b j_b^{(n)\mu}=0, 
  \label{conservation_small_j}
\end{eqnarray}
because if these relations hold, then 
(\ref{EOM:partial_n-th_order}) and 
(\ref{EOM:O_n-th_order}) can be solved as in the case of 
(\ref{eq:EOM_partial_lowest_order}). 

First let us show (\ref{conservation_large_J}). 
From (\ref{field strength}),   
$\partial^a\partial^b F_{ab}^{(n)cd}$ vanish identically.  
Therefore, we have 
\begin{eqnarray}
  \partial^b J_{b}^{(n)cd}
  =
  -\partial^b
  \left(
    \partial^a F_{ab}^{(n)cd}
    -
    J_{b}^{(n)cd}
  \right). 
\end{eqnarray}
By definition, the right hand side of this equation is 
equal to 
\begin{eqnarray}
  -\partial^b
  \left(
    \left.
      \left[
        D^a,\left[D_a,D_b\right]
      \right]^{(n)}
    \right|_{{\cal O}}
  \right)^{cd}. 
\end{eqnarray}
Noticing that 
\begin{eqnarray}
  \lefteqn{ 
    \left[\partial^b,\left[D^a,\left[D_a,D_b\right]\right]^{(n)}\right]
  }\nonumber\\
  &=&
  \partial^b\left(
    \left.
      \left[
        D^a,\left[D_a,D_b\right]
      \right]^{(n)}
    \right|_\partial
  \right)^\mu\cdot\partial_\mu
  \nonumber\\
  & &
  +
  \partial^b\left(
    \left.
      \left[
        D^a,\left[D_a,D_b\right]
      \right]^{(n)}
    \right|_{{\cal O}}
  \right)^{cd}\cdot{\cal O}_{cd}
  -
  \left(
    \left.
      \left[
        D^a,\left[D_a,D_b\right]
      \right]^{(n)}
    \right|_{{\cal O}}
  \right)_b{}^{c}\partial_c,  
  \label{eq:current_conservation_and_commutator}
\end{eqnarray}
we obtain 
\begin{eqnarray}
  -\partial^b
  \left(
    \left.
      \left[
        D^a,\left[D_a,D_b\right]
      \right]^{(n)}
    \right|_{{\cal O}}
  \right)^{cd}
  =
  \left(
    \left.
      -\left[\partial^b,\left[D^a,\left[D_a,D_b\right]\right]^{(n)}\right]
    \right|_{{\cal O}}
  \right)^{cd}. 
\end{eqnarray}
Then, using the equations of motion in the lower orders 
(\ref{EOM:lower order}) we have  
\begin{eqnarray} 
  \left(
    \left.
      -\left[\partial^b,\left[D^a,\left[D_a,D_b\right]\right]^{(n)}\right]
    \right|_{{\cal O}}
  \right)^{cd}
  =
  \left(
    \left.
      -\left[D^b,\left[D^a,\left[D_a,D_b\right]\right]\right]^{(n)}
    \right|_{{\cal O}}
  \right)^{cd}. 
  \label{triple_commutator_O}
\end{eqnarray}
Finally, from the identity 
$\left[D^b,\left[D^a,\left[D_a,D_b\right]\right]\right]=0$, 
which is easily derived from the Jacobi identity, 
the right hand side of (\ref{triple_commutator_O}) is zero. 
This completes the proof of (\ref{conservation_large_J}), 
and hence we can find a solution of (\ref{EOM:O_n-th_order})
for $S_a^{\prime (n)cd}$. 

Next we show (\ref{conservation_small_j}). 
In this case  
we must use the equation of motion for $S_a{}^{\prime(n)bc}$, 
in addition to ones for $h^{(i)}$ and $S^{\prime(i)}(i=1,\cdots,n-1)$, 
because $j_b^{(n)\mu}$ depends on $S^{\prime(n)}$  
as can be seen from (\ref{eq:small_j}). 
As before, using (\ref{field strength}) and 
(\ref{eq:current_conservation_and_commutator})  
we have 
\begin{eqnarray}
  \partial^b j_{b}^{(n)\mu}
  &=&
  -\partial^b
  \left(
    \partial^a f_{ab}^{(n)\mu}
    -
    j_b^{(n)\mu}
  \right)
  \nonumber\\
  &=&
  -\partial^b
  \left(
    \left.
      \left[D^a,\left[D_a,D_b\right]\right]^{(n)}
    \right|_{\partial}
  \right)^{cd}
  \nonumber\\
  &=&     
  \left(
    \left.
      -\left[\partial^b,\left[D^a,\left[D_a,D_b\right]\right]^{(n)}\right]
    \right|_{\partial}
  \right)^\mu
  -
  \left(
    \left.
      \left[D^a,\left[D_a,D_b\right]\right]^{(n)}
    \right|_{\cal O}
  \right)_b{}^\mu. 
  \label{conservation_Jh}
\end{eqnarray} 
Then, by using (\ref{EOM:O_n-th_order}) and 
the identity 
$\left[D^b,\left[D^a,\left[D_a,D_b\right]\right]\right]=0$ 
the last expression becomes zero, and hence we have 
(\ref{conservation_small_j}). 

In this way, we can construct the solution order by order. 
Note that the resemblance to Yang-Mills theory is crucial for 
the existence of the solution.   
In appendix \ref{appendix:DOF}, we discuss how many physical degrees 
of freedom remain after taking the gauge symmetries into account.    
%%%%%%%%%%%%%%%%%%%%%%%%%%%%%%%%%%%%%%%%%%%%%%%%%%%%%%%%%%
%%%%%%%%%%%%%%%%%%%%%%%%%%%%%%%%%%%%%%%%%%%%%%%%%%%%%%%%%%
%%%%%%%%%%%%%%%%%%%%%%%%%%%%%%%%%%%%%%%%%%%%%%%%%%%%%%%%%%
\subsection{Further Analysis for $\phi$ }
%\label{subsec:difficulty_of_L=0}
%%%%%%%%%%%%%%%%%%%%%%%%%%%%%%%%%%%%%%%%%%%%%%%%%%%%%%%%%%
%%%%%%%%%%%%%%%%%%%%%%%%%%%%%%%%%%%%%%%%%%%%%%%%%%%%%%%%%%
%%%%%%%%%%%%%%%%%%%%%%%%%%%%%%%%%%%%%%%%%%%%%%%%%%%%%%%%%%
\noindent
\hspace{4.6mm}
In section \ref{subsec:linear}, we showed that at the linearized level 
the equations of motion (\ref{nabla-eq R T}) 
and (\ref{O-eq R T}) have a solution composed of 
the scalar $\phi$ and the rank-two tensor $B_{ab}$. 
In section \ref{The existence of solutions} we showed that 
any solution at the linearized level 
can be lifted to a solution of full orders 
in the weak field expansion. 
Although the degrees of freedom other than $\phi$ and $B_{ab}$ 
are not desirable as propagating fields, they are necessary 
to make full-order solutions.  
In order to clarify this point, 
we discuss the difficulty of the following ansatz:  
\begin{eqnarray}
  V_a=\nabla_a\phi, 
  \qquad
  H_{abc}=0, 
  \qquad
  L_{abc}=0.    
  \label{restricted_ansatz}
\end{eqnarray}
We show that (\ref{restricted_ansatz}) gives a rather strong condition 
on $\phi$ so that $\phi$ cannot be regarded as a scalar field. 
Note that we impose this condition {\it to all orders}. 
In this ansatz, the Einstein equation 
(\ref{nabla-eq R T}) becomes 
\begin{eqnarray}
  R_{ab}
  =
  (3-d)\nabla_a\nabla_b\phi
  +
  (2-d)(\nabla_a\phi)(\nabla_b\phi)
  -
  2(\nabla\cdot\phi)\delta_{ab}
  +
  (2d-3)(\nabla\phi)^2\delta_{ab}. 
  \label{eq:Einstein_with_only_phi}
\end{eqnarray}
Substituting it to the Bianchi identity
\begin{eqnarray}
  2\nabla^a R_{ab}
  -
  \nabla_b R
  =
  0, 
\end{eqnarray}
we obtain 
\begin{eqnarray}
  \lefteqn{
    \partial_a\left\{
      (d-1)\nabla^2\phi
      +
      (-d^2+d+3)(\nabla\phi)^2
    \right\}
  }\nonumber\\
  & &
  +
  (\partial_a\phi)\left\{
    (2d-8)(\nabla^2\phi)
    +
    2(3-d)(d-1)(\nabla\phi)^2
  \right\}
  =0. 
  \label{eq:Bianchi_for_dilaton_2}
\end{eqnarray}
This condition is too stringent for $\phi$ to be a scalar field. 
Indeed, in the weak field expansion, a generic solution 
of the first order cannot be lifted to all orders. 

At the first order (\ref{eq:Bianchi_for_dilaton_2}) reduces to 
\begin{eqnarray}
  \partial^2\phi^{(1)}=0,  
  \label{Laplace_eq_for_phi1}
\end{eqnarray}
and hence $\phi^{(1)}$ is a harmonic function. 
Then, we determine $e_a^{(1)\mu}$ from 
(\ref{eq:Einstein_with_only_phi}). 
Using (\ref{Laplace_eq_for_phi1}), $\nabla^2\phi$ becomes a quantity 
of the second order, and hence 
the second term in the l.~h.~s. of (\ref{eq:Bianchi_for_dilaton_2}) 
is a quantity of the third order. 
Therefore, at the second order
 (\ref{eq:Bianchi_for_dilaton_2}) becomes
\begin{eqnarray}
  \left\{
    (d-1)\nabla^2\phi+(-d^2+d+3)(\nabla\phi)^2
  \right\}^{(2)}
  =
  c, 
  \label{second_order}
\end{eqnarray}
where $c$ is a constant. 
Because $(\nabla^2\phi)^{(2)}$ can be rewritten as 
\begin{eqnarray}
  (\nabla^2\phi)^{(2)}
  =
  \partial^2\phi^{(2)}
  +
  (\nabla^2\phi^{(1)})^{(2)}, 
\end{eqnarray}
we can solve $\phi^{(2)}$ in terms of $\phi^{(1)}$ 
and $e_a^{(1)\mu}$, and then we can determine $e_a^{(2)\mu}$ 
from (\ref{eq:Einstein_with_only_phi}). 
At the third order, (\ref{eq:Bianchi_for_dilaton_2}) becomes 
\begin{eqnarray}
  0
  &=&
  \partial_a\left\{
    (d-1)\partial^2\phi^{(3)}
    +
    \left(
      \mbox{polynomial of }\phi^{(1)},\phi^{(2)},e^{(1)},e^{(2)}
    \right)
  \right\}
  \nonumber\\
  & &
  +
  \left(
    \partial_a\phi^{(1)}
  \right)
  \left\{
    (2d-8)\nabla^2\phi
    +
    2(3-d)(d-1)
    (\partial\phi)^2
  \right\}^{(2)}
  \nonumber\\
  &=&
   \partial_a\left\{
      (d-1)\partial^2\phi^{(3)}
      +
      \left(
        \mbox{polynomial of }\phi^{(1)},\phi^{(2)},e^{(1)},e^{(2)}
      \right)
    \right\}
  \nonumber\\
  & &
  +
  \left(
    \partial_a\phi^{(1)}
  \right)
  \left\{
    (2d-8)\frac{d^2-d-3}{d-1}
    +
    2(3-d)(d-1)
  \right\}
  \left(
    \partial\phi^{(1)}
  \right)^2
  \nonumber\\
  & &
  +
  c\cdot\frac{2d-8}{d-1}\partial_a\phi^{(1)},  
  \label{eq:third_order}
\end{eqnarray}
where we have used (\ref{second_order}) at the second equality. 
In order for $\phi^{(3)}$ to exist, 
the second term of the last expression  
must satisfy the integrability condition, that is,  
$
\left(
  \partial_a\phi^{(1)}
\right)
\left(
  \partial\phi^{(1)}
\right)^2 
$ must be total derivative. 
However, this is not the case 
for generic $\phi^{(1)}$. 
Therefore, we cannot set $H_{abc}=0$ and $L_{abc}=0$
%f1 as in (\ref{restricted_ansatz})
; even if we impose $H^{(1)}_{abc}=L^{(1)}_{abc}=0$, 
$H_{abc}$ and $L_{abc}$ appears in higher orders 
as nonlocal quantities obtained from $\phi$.  
%%%%%%%%%%%%%%%%%%%%%%%%%%%%%%%%%%%%%%%%%%%%%%%%%%%%%%%%%%
%%%%%%%%%%%%%%%%%%%%%%%%%%%%%%%%%%%%%%%%%%%%%%%%%%%%%%%%%%
%%%%%%%%%%%%%%%%%%%%%%%%%%%%%%%%%%%%%%%%%%%%%%%%%%%%%%%%%%
\section{Matrix Model with Additional Terms}\label{sec:modified_action}
%%%%%%%%%%%%%%%%%%%%%%%%%%%%%%%%%%%%%%%%%%%%%%%%%%%%%%%%%%
%%%%%%%%%%%%%%%%%%%%%%%%%%%%%%%%%%%%%%%%%%%%%%%%%%%%%%%%%%
%%%%%%%%%%%%%%%%%%%%%%%%%%%%%%%%%%%%%%%%%%%%%%%%%%%%%%%%%%
\hspace{4.6mm}
In this section, we study matrix models with a mass term 
and a cubic term, which were introduced 
in the context of 
D-brane action in background fields 
\cite{Myers:1999ps,BMN}
or as toy models to obtain nontrivial vacua \cite{IKTW,ABN}.
We analyze the effects of these terms in our interpretation. 
In section \ref{subsec:action_with_mass_term}, 
we consider matrix models with a mass term.  
In the paper \cite{HKK}, 
it was shown that 
Einstein manifolds are obtained as classical solutions 
of these models, and for such solutions 
the coefficient of the mass term plays the role of 
the cosmological constant. 
If we introduce a torsion, 
we find another type of solutions that have constant background 
of field strength of $B$. 
In this case, the coefficient of the mass term is given by the 
field strength. 
In section \ref{subsec:action_with_cubic_term}, 
we consider a three-dimensional matrix model with a cubic term.  
We find that this model also has the second type of  
classical solution in which  
the coefficient of the cubic term can be interpreted as 
the field strength. 
In both cases, the antisymmetric part of the torsion 
equals the field strength of $B$.  
This result supports the interpretation  
that the $B$-field is embedded in the torsion. 
%%%%%%%%%%%%%%%%%%%%%%%%%%%%%%%%%%%%%%%%%%%%%%%%%%%%%%%%%%
%%%%%%%%%%%%%%%%%%%%%%%%%%%%%%%%%%%%%%%%%%%%%%%%%%%%%%%%%%
%%%%%%%%%%%%%%%%%%%%%%%%%%%%%%%%%%%%%%%%%%%%%%%%%%%%%%%%%%
\subsection{Matrix Model with Mass 
Term}\label{subsec:action_with_mass_term}
%%%%%%%%%%%%%%%%%%%%%%%%%%%%%%%%%%%%%%%%%%%%%%%%%%%%%%%%%%
%%%%%%%%%%%%%%%%%%%%%%%%%%%%%%%%%%%%%%%%%%%%%%%%%%%%%%%%%%
%%%%%%%%%%%%%%%%%%%%%%%%%%%%%%%%%%%%%%%%%%%%%%%%%%%%%%%%%%
\hspace{4.6mm}
In this subsection, we
consider the matrix model with a mass term
\begin{eqnarray}
  S=-\frac{1}{4g^2}Tr[A_a,A_b][A^a,A^b]
  -\frac{\alpha^2}{2g^2}\delta_{ab}Tr\left(A^aA^b\right), 
  \label{action_with_mass_term}  
\end{eqnarray}
where $\alpha$ is a parameter with the dimension of mass. 
We show that this model has 
two types of solutions without and with torsion. 
The former consists of the Einstein manifolds and 
the latter does group manifolds with constant 
field strength of $B$. 

First, let us consider the case where the torsion is absent 
\cite{HKK}. 
For an ansatz $A_{a}=i\nabla_{(a)}$, 
the equation of motion 
\begin{eqnarray}
  \left[A^a,[A_a,A_b]\right]
  -
  \alpha^2 A_b
  =
  0 
  \label{eqationofmotionwithmassterm}
\end{eqnarray}
becomes 
\begin{eqnarray}
  \left[\nabla^a,[\nabla_a,\nabla_b]\right]
  +
  \alpha^2  \nabla_b
  =
  -\left(R_b{}^c-\alpha^2 \delta_b{}^c\right)\nabla_c
    +
    (\nabla^a R_{ab}{}^{cd}){\cal O}_{cd}
  =
  0, 
\end{eqnarray}
and this is satisfied if the spacetime is an Einstein manifold, 
\begin{eqnarray}
  R_b{}^c=\alpha^2 \delta_b{}^c.  
  \label{sol_with_cc}
\end{eqnarray}
The quantity $\alpha^{2}$ 
plays the role of 
a cosmological constant. 

We next consider the case in which the torsion is present. 
For an ansatz $A_{a}=iD_{(a)}$, where $D_{a}$ is a covariant 
derivative {\it with} a torsion, 
the equation of motion (\ref{eqationofmotionwithmassterm}) 
again becomes 
\begin{eqnarray}
  \left[D^a,[D_a,D_b]\right]
  +\alpha^2 D_b=0.  
  \label{eom_in_terms_of_D}
\end{eqnarray}
If the operator $D_a$ is proportional to 
the standard left-invariant derivative 
on a Lie group $G$, they satisfy the commutation relation 
of the Lie algebra, 
\begin{eqnarray}
  [D_a, D_b]
  =
  \alpha f_{abc}D_{c}, 
  \label{ansatz:commutation_relation}
\end{eqnarray}
where $f_{abc}$ is the structure constant of $G$.  
Then the equation of motion (\ref{eom_in_terms_of_D}) 
is satisfied if $f_{abc}$ is normalized as  
\begin{eqnarray}
  f_{acd}f_b{}^{cd}=\delta_{ab}. 
\end{eqnarray}

As shown in appendix 
\ref{appendix:properties_of_group_manifold}, 
we can interpret $D_a$ 
as the covariant derivative on $G$ with torsion, 
\begin{eqnarray}
  D_a
  =
  \nabla_a
  +
  S_a{}^{bc}{\cal O}_{bc},
  \qquad
  S_{abc}
  =
  H_{abc}
  =
  \frac{\alpha}{2}f_{abc}. 
  \label{eq:torsion_and_structure_constant}
\end{eqnarray}
Here, the vielbein is defined by 
\begin{eqnarray}
  t_a
  =
  e_a{}^\mu\cdot i\alpha g^{-1}\partial_\mu g, 
\end{eqnarray}
where $g(x)$ is an element of $G$, $x$ is a coordinate on $G$ 
and $t_a$ is a generator of $G$, 
and we have 
\begin{eqnarray}
  & &
  R_{ab}{}^{cd}=\frac{\alpha^2}{4}f_{abe}f^{cde}
  =
  H_{abe}H^{cde}, 
  \label{Riemann_in_terms_of_structure_constant}
  \\
  & &
  \nabla_a S_{bcd}=
  \nabla_a H_{bcd}=0. 
  \label{eq:H is covariantly constant}
\end{eqnarray}
The torsion (\ref{eq:torsion_and_structure_constant}) 
can be regarded as a constant field strength of $B_{ab}$,  
\begin{eqnarray}
  H_{abc}=\nabla_{[a}B_{bc]}.  
\end{eqnarray}
Indeed, from (\ref{Riemann_in_terms_of_structure_constant}) and 
(\ref{eq:H is covariantly constant}) 
we obtain the equations 
\begin{eqnarray}
  &&R_{a}{}^{c}=H_{aef}H^{cef}, 
  \label{einstein}\\ 
  &&\nabla^a H_{abc}=0,
  \label{eom for B}
\end{eqnarray}
which are nothing but the equations of motion 
for the metric tensor and the $B$-field
\footnote{
  %If we also consider the dilaton $\phi$, its equation of motion 
  %gives 
  %$H_{abc}H^{abc}=0$ if we simply set $\phi$ to be a constant and 
  %the cosmological constant to be zero.  
  %However,     
  %(\ref{einstein}) and (\ref{eom for B}) are consistent with 
  %the equations of motions for the system consists of 
  %the graviton, the $B$-field and the dilaton 
  %with a cosmological constant $\Lambda=\frac{d}{12}\alpha^2$. 
  This set of equations can also be regarded as equations for 
  the graviton, the $B$-field and a constant dilaton with 
  a cosmological constant $\Lambda=\frac{d}{12}\alpha^2$. 
}.  
Note that we could set $V_a=0$ and $L_{abc}=0$ in this case 
while it is not the case when the mass term is absent. 
%%%%%%%%%%%%%%%%%%%%%%%%%%%%%%%%%%%%%%%%%%%%%%%%%%%%%%%%%%
%%%%%%%%%%%%%%%%%%%%%%%%%%%%%%%%%%%%%%%%%%%%%%%%%%%%%%%%%%
%%%%%%%%%%%%%%%%%%%%%%%%%%%%%%%%%%%%%%%%%%%%%%%%%%%%%%%%%%
\subsection{Matrix Model with Cubic 
Term}\label{subsec:action_with_cubic_term}
%%%%%%%%%%%%%%%%%%%%%%%%%%%%%%%%%%%%%%%%%%%%%%%%%%%%%%%%%%
%%%%%%%%%%%%%%%%%%%%%%%%%%%%%%%%%%%%%%%%%%%%%%%%%%%%%%%%%%
%%%%%%%%%%%%%%%%%%%%%%%%%%%%%%%%%%%%%%%%%%%%%%%%%%%%%%%%%%
\hspace{4.6mm}
In this subsection, we show that the three-dimensional 
matrix model with a cubic term 
also has the group manifold $SU(2)$ as a classical solution. 

Let us consider the action 
\begin{eqnarray}
  S
  =
  -
  \frac{1}{4g^2}Tr[A_a,A_b][A^a,A^b]
  +
  \frac{i\alpha}{3g^2}\epsilon_{abc}Tr[A^a,A^b]A^c,  
  \label{action_with_cubic_term}
\end{eqnarray}
where $\epsilon_{abc}$ is the structure constant of 
$SU(2)$. 
For an ansatz $A_a=iD_{(a)}$, the equation of motion is 
\begin{eqnarray}
  \left[
    D_{(a)},[D_{(a)},D_{(b)}]
  \right]
  -
  \alpha\epsilon_{bcd}[D_{(c)},D_{(d)}]
  =
  0. 
  \label{EOM_with_parentheses}
\end{eqnarray}
Because the structure constant $\epsilon_{abc}$ is an invariant 
tensor of the Lorentz group $Spin(3)=SU(2)$, we can rewrite 
(\ref{EOM_with_parentheses}) as 
\begin{eqnarray}
  \left[
    D_{a},[D_{a},D_{b}]
  \right]
  -
  \alpha\epsilon_{bcd}[D_{c},D_{d}]
  =
  0. 
\end{eqnarray}
Therefore, if operators $D_{a}$ satisfy the commutation relation 
\begin{eqnarray}
  [D_{a}, D_{b}]
  =
  \alpha \epsilon_{abc}D_{c},  
  \label{commutation_relation_without_parenthesis_SU(2)}
\end{eqnarray}
then $A_a=iD_{(a)}$ is a classical solution. 
As in the case of the previous subsection, this solution 
describes the group manifold $SU(2)\simeq S^3$ with a constant 
field strength of $B$. 

\vspace{0.4cm}

We now comment on dynamical properties of 
classical solutions. 
The solution $A_a=iD_{(a)}$ 
% , which 
% is the covariant derivative with the torsion 
% on a group manifold, 
can be expressed 
by a Lie algebra corresponding to the group manifold. 
Therefore 
we can discuss dynamical properties of this background 
using techniques of the noncommutative geometry. 
On the other hand, the solution without the flux does not form 
a Lie algebra, and hence the analysis is rather difficult. 
As a concrete example, we discuss a stability of $S^{3}$ 
in appendix \ref{appendix:stability_instability}. 
%%%%%%%%%%%%%%%%%%%%%%%%%%%%%%%%%%%%%%%%%%%%%%%%%%%%%%%%%%
%%%%%%%%%%%%%%%%%%%%%%%%%%%%%%%%%%%%%%%%%%%%%%%%%%%%%%%%%%
%%%%%%%%%%%%%%%%%%%%%%%%%%%%%%%%%%%%%%%%%%%%%%%%%%%%%%%%%%
\section{Conclusions and Discussions}\label{sec:Discussions}
%%%%%%%%%%%%%%%%%%%%%%%%%%%%%%%%%%%%%%%%%%%%%%%%%%%%%%%%%%
%%%%%%%%%%%%%%%%%%%%%%%%%%%%%%%%%%%%%%%%%%%%%%%%%%%%%%%%%%
%%%%%%%%%%%%%%%%%%%%%%%%%%%%%%%%%%%%%%%%%%%%%%%%%%%%%%%%%%
\hspace{4.6mm}
In the previous paper, we have introduced a new interpretation 
of the matrix model where matrices act on 
the space of functions on the principal $Spin(d)$ bundle over 
a Riemannian manifold. 
In this paper, we studied the equation of motion for 
the vielbein and torsion in the matrix model. 
In section \ref{sec:Torsion_as_massless_fields}, 
we interpreted the trace and antisymmetric part of the torsion 
to be field strengths of the dilaton and the $B$-field, respectively. 
We found a solution in which only  
the graviton, dilaton and $B$-field propagate. 
In general, however, all components of the torsion, 
including the negative-norm modes, can propagate, 
and  there is no gauge symmetry to remove them. 
In spite of this difficulty, 
there are several possibilities that  
these unnecessary fields disappear from the spectrum.  
First, recall that higher spin fields couple to the torsion fields 
in a complicated manner, although we have simply set them to zero. 
They may give the large mass for the unnecessary degrees of 
freedom through the interactions. 
Also there is a possibility that 
the mass terms of unnecessary fields 
arise through quantum corrections. 
If the unnecessary fields disappear through such mechanisms, 
then our system can reproduce the massless bosonic sector of 
superstring theory.  
In section \ref{sec:modified_action}, 
we studied the matrix model with mass and cubic terms, 
and found solutions with a constant torsion. In this case, 
the antisymmetric part of the torsion can be interpreted 
as the constant background of field strength of $B$. 
This result adds another ground to regard 
the antisymmetric part of the torsion 
to be the field strength of the $B$-field. 

If unnecessary fields do not disappear from the spectrum dynamically,  
we may need to impose certain constraints to eliminate them. 
Although to find a good constraint seems difficult, 
it might become easier if we first extend the matrix space. 
One possible way to extend the matrix space is to replace 
usual manifolds with supermanifolds \cite{HKK2}. 
This corresponds to consider 
supermatrices instead of ordinary matrices. 
Due to this extension, we can implement 
the local supersymmetry as a subset of the unitary symmetry. 
In this case,  
a supermultiplet including the gravity is embedded successfully 
and the torsion constraint removes 
unnecessary components of the torsion. 
In particular, in the case of four dimensions 
it was shown that 
the equation of motion of the supermatrix model 
is consistent with that of ${\cal N}=1$ supergravity. 
This result suggests that 
a good constraint such as the torsion constraint 
can be found by extending the matrix space.  
Although in ten dimensions the superspace formulation 
is rather complicated and it is not clear whether 
the torsion constraint fits well to our scheme, 
the research along this line would be helpful 
to understand how the spectrum of 
string theory is derived from 
the matrix model. 
%%%%%%%%%%%%%%%%%%%%%%%%%%%%%%%%%%%%%%%%%%%%%%%%%%%%%%%%%%
%%%%%%%%%%%%%%%%%%%%%%%%%%%%%%%%%%%%%%%%%%%%%%%%%%%%%%%%%%
%%%%%%%%%%%%%%%%%%%%%%%%%%%%%%%%%%%%%%%%%%%%%%%%%%%%%%%%%%
\vspace{0.3cm}
\begin{center} \begin{large}
Acknowledgments
\end{large} \end{center}
%%%%%%%%%%%%%%%%%%%%%%%%%%%%%%%%%%%%%%%%%%%%%%%%%%%%%%%%%%
%%%%%%%%%%%%%%%%%%%%%%%%%%%%%%%%%%%%%%%%%%%%%%%%%%%%%%%%%%
%%%%%%%%%%%%%%%%%%%%%%%%%%%%%%%%%%%%%%%%%%%%%%%%%%%%%%%%%%
M.~H. thanks T.~Azeyanagi, T.~Azuma and T.~Hirata 
for stimulating discussions and comments.  
Y.~K. is grateful to  F.~Sugino for helpful  discussions.   
%He also expresses his sincere thanks to {\it Deep Impact}. 
%His effort toward the victory of Prix de l'Arc de Triomphe 
%encourages Y.~K. continuously.  
M.~H. and Y.~K. also 
thank the Japan Society for the Promotion of Science for financial
support. 
K.~F. and M.~H. were in part supported by JSPS and French Ministry of 
Foreign Affairs under the Japan-France Integrated Action 
Program (SAKURA). 
M.~H. would like to thank CEA/Saclay and Ivan Kostov for hospitality 
while part of this work was done. 
This work was also supported in part by a Grant-in-Aid for
the 21st Century COE ``Center for Diversity and Universality in
Physics''.

%%%%%%%%%%%%%%%%
\appendix
%%%%%%%%%%%%%%%%%%%%%%%%%%%%%%%%%%%%%%%%%%%%%%%%%%%%%%%%%%
%%%%%%%%%%%%%%%%%%%%%%%%%%%%%%%%%%%%%%%%%%%%%%%%%%%%%%%%%%
%%%%%%%%%%%%%%%%%%%%%%%%%%%%%%%%%%%%%%%%%%%%%%%%%%%%%%%%%%
\section{Physical Degrees of Freedom}\label{appendix:DOF}
%%%%%%%%%%%%%%%%%%%%%%%%%%%%%%%%%%%%%%%%%%%%%%%%%%%%%%%%%%
%%%%%%%%%%%%%%%%%%%%%%%%%%%%%%%%%%%%%%%%%%%%%%%%%%%%%%%%%%
%%%%%%%%%%%%%%%%%%%%%%%%%%%%%%%%%%%%%%%%%%%%%%%%%%%%%%%%%%
\hspace{4.6mm}
In this section, we consider the physical degrees of freedom of 
the torsion.  
We can impose the gauge fixing condition  
\begin{eqnarray}
  \partial^a S^\prime_a{}^{bc}=0, 
  \qquad
  \partial^a h_a{}^\mu=0 
\end{eqnarray}
at all orders in the weak field expansion. 
This condition removes the components along the light-cone direction, 
$h_+{}^\mu$ and $S^\prime_+{}^{cd}$.  
Furthermore, because the equation of motion 
(\ref{proof_of_existence_of_solution_1}) and 
(\ref{proof_of_existence_of_solution_2}) can be regarded as 
the equation of motion for Yang-Mills theory with extra indices 
$\mu,c$ and $d$, we can use the residual symmetries to remove 
$h_-{}^\mu$ and $S^\prime_-{}^{cd}$. Therefore, the physical degrees of 
freedom are  
\begin{eqnarray}
  h_i{}^\mu, 
  \qquad
  S^\prime_i{}^{cd}. 
\end{eqnarray}
Here, $i$ represents the transverse direction, $i=2,3,\cdots,d-1$. 
In terms of irreducible tensors with respect to the transverse 
$SO(d-2)$, they can be decomposed as follows:
\begin{eqnarray}
  & &
  h_{i}{}^0, h_{i}{}^1, h_i{}^i, h_{[ij]}, 
  h_{(ij)}-\frac{h_{(kk)}}{d-2}\delta_{ij}, 
  \nonumber\\
  & & 
  S^\prime_i{}^{01}, S^\prime_i{}^{i0}, S^\prime_i{}^{i1}, 
  S^\prime_i{}^{ij}, 
%f2
  S^\prime_{(ij)}{}^0-\frac{S^\prime_{(kk)}{}^0 \delta_{ij}}{d-2}, 
  S^\prime_{(ij)}{}^1-\frac{S^\prime_{(kk)}{}^1 \delta_{ij}}{d-2}, 
  \nonumber\\
  & &
  S^\prime_{[ij]}{}^0, S^\prime_{[ij]}{}^1,  
  S^\prime_{[ijk]}, 
  S^\prime_{i}{}^{jk}
  -
  \frac{S^\prime_l{}^{lk}\delta_i{}^j}{d-2}
  -
  S^\prime_{[ijk]}
\end{eqnarray}
Unlike in the Yang-Mills theory, fields with indices
$0$ and $1$ are not completely removed.
Quantities including $0$ as indices,  
$h_i{}^0,S^\prime_i{}^{0j}$ and $S^\prime_i{}^{01}$, 
are negative-norm modes. 
%%%%%%%%%%%%%%%%%%%%%%%%%%%%%%%%%%%%%%%%%%%%%%%%%%%%%%%%%%
%%%%%%%%%%%%%%%%%%%%%%%%%%%%%%%%%%%%%%%%%%%%%%%%%%%%%%%%%%
%%%%%%%%%%%%%%%%%%%%%%%%%%%%%%%%%%%%%%%%%%%%%%%%%%%%%%%%%%
\section{Differential Calculus on Lie 
Group}\label{appendix:properties_of_group_manifold}
%%%%%%%%%%%%%%%%%%%%%%%%%%%%%%%%%%%%%%%%%%%%%%%%%%%%%%%%%%
%%%%%%%%%%%%%%%%%%%%%%%%%%%%%%%%%%%%%%%%%%%%%%%%%%%%%%%%%%
%%%%%%%%%%%%%%%%%%%%%%%%%%%%%%%%%%%%%%%%%%%%%%%%%%%%%%%%%%
\hspace{4.6mm}
In this section, we provide basic results concerning the differential 
calculus on a Lie group $G$, which are necessary in 
section \ref{sec:modified_action}. 
For further details and other applications to physics, 
see \cite{BCZ} for example. 
%%%%%%%%%%%%%%%%%%%%%%%%%%%%%%%%%%%%%%%%%%%%%%%%%%%%%%%%%%
%%%%%%%%%%%%%%%%%%%%%%%%%%%%%%%%%%%%%%%%%%%%%%%%%%%%%%%%%%
%%%%%%%%%%%%%%%%%%%%%%%%%%%%%%%%%%%%%%%%%%%%%%%%%%%%%%%%%%
%\subsection{Metric, Christoffel Symbol and Spin Connection}
%%%%%%%%%%%%%%%%%%%%%%%%%%%%%%%%%%%%%%%%%%%%%%%%%%%%%%%%%%
%%%%%%%%%%%%%%%%%%%%%%%%%%%%%%%%%%%%%%%%%%%%%%%%%%%%%%%%%%
%%%%%%%%%%%%%%%%%%%%%%%%%%%%%%%%%%%%%%%%%%%%%%%%%%%%%%%%%%

Let $t_a$ be the generator of a Lie group 
$G$ which satisfies the relations 
\begin{eqnarray}
  [t_a,t_b]=if_{abc}t_c. 
  \label{commutation_relation_of_t}
\end{eqnarray}
The left-invariant derivative $D_a$ on $G$ is defined by 
\begin{eqnarray}
  \epsilon^a D_a f(g)
  =
  f(g(1-i\epsilon^a t_a))
  -
  f(g), 
  \label{definition of left inv deriv}
\end{eqnarray}
where $f(g)$ is a function on $G$. 
From (\ref{commutation_relation_of_t}), we can easily see 
that $D_a$ satisfies the commutation relation   
\begin{eqnarray}
  [D_a,D_b]=f_{abc}D_c. 
  \label{commutation_relation_of_D}
\end{eqnarray}
By introducing a local coordinate $x$, $D_a$ can be expressed as 
\begin{eqnarray}
  D_a=v_a{}^\mu\partial_\mu, 
  \label{vielbein_1}
\end{eqnarray}
where $\partial_\mu=\partial /\partial x^\mu$ and 
$v_a{}^\mu(x)\ (a=1,\cdots,\mbox{dim} G)$ are vector fields on $G$. 
By taking $f(g)=g$, $v_a{}^\mu$ can be expressed as 
\begin{eqnarray}
  t_a
  =
  v_a{}^\mu\cdot ig^{-1}(x)\partial_\mu g(x),
  \label{vielbein_2} 
\end{eqnarray}
where $g(x)$ is the element of $G$ specified by the coordinate $x$. 

Next we identify $v_a{}^\mu$ to a vielbein $e_a{}^\mu$.  
From this vielbein, we obtain the standard covariant derivative 
$\nabla_a=
e_a{}^\mu
\left(
\partial_\mu+\omega_\mu{}^{bc}{\cal O}_{bc}
\right)$, 
where $\omega$ is the spin connection.  
From (\ref{vielbein_1}),  
$\nabla_a$ is related to $D_a$ as  
\begin{eqnarray}
  D_a=\nabla_a-\omega_a{}^{bc}{\cal O}_{bc}. 
  \label{relationship_D_and_nabla}
\end{eqnarray}
Then, if we regard the second term of the r.h.s. of 
(\ref{relationship_D_and_nabla}) as contorsion, 
$D_a$ can be regarded as the covariant 
derivative with torsion, 
\begin{eqnarray}
  D_a=\nabla_a+S_a{}^{bc}{\cal O}_{bc}, 
  \qquad
  S_{abc}=-\omega_{abc}. 
\end{eqnarray}
Combining it with (\ref{commutation_relation_of_D}), 
we have 
\begin{eqnarray}
  & &
  {\cal R}_{ab}{}^{cd}=0, 
  \label{vanishing_of_cov_deriv_with_torsion}
  \\
  & &
  S_{abc}=-\omega_{abc}=\frac{1}{2}f_{abc}. 
\end{eqnarray}

We can show the relation $\nabla_a f_{bcd}=0$ 
by using the Jacobi identity:  
\begin{eqnarray}
  2\nabla_{a}S_{bcd}
  &=&
  \nabla_a f_{bcd}
  \nonumber\\
  &=&
  e_a{}^\mu\partial_\mu f_{bcd}
  +
  \omega_{ab}{}^e f_{ecd}
  +
  \omega_{ac}{}^e f_{bed}
  +
  \omega_{ad}{}^e f_{bce}
  \nonumber\\
  &=&
  -\frac{1}{2}\left(
    f_{ab}{}^e f_{ecd}
    +
    f_{ac}{}^e f_{bed}
    +
    f_{ad}{}^e f_{bce}  
  \right)
  \nonumber\\
  &=&
  0. 
  \label{eq:cov_deriv_of_str_const}
\end{eqnarray}
By using (\ref{definition of cal R and T}), 
(\ref{Riemann_tensor_with_torsion}), 
(\ref{vanishing_of_cov_deriv_with_torsion})   
and (\ref{eq:cov_deriv_of_str_const}),   
the Riemann tensor can be expressed in terms of 
the structure constant (or equivalently, the torsion) as  
\begin{eqnarray}
  R_{ab}{}^{cd}
  &=&
  -
  \nabla_a S_b{}^{cd}
  +
  \nabla_b S_a{}^{cd}
  +
  S_{ae}{}^c S_b{}^{ed}
  -
  S_{be}{}^c S_a{}^{ed}
  \nonumber\\
  &=&
  \frac{1}{4}\left(
  f_{ae}{}^cf_b{}^{ed}
  -
  f_{be}{}^cf_a{}^{ed}
  \right)
  \nonumber\\
  &=&
  \frac{1}{4}
  f_{abe}f^{cde}. 
\end{eqnarray}
%%%%%%%%%%%%%%%%%%%%%%%%%%%%%%%%%%%%%%%%%%%%%%%%%%%%%%%%%%
%%%%%%%%%%%%%%%%%%%%%%%%%%%%%%%%%%%%%%%%%%%%%%%%%%%%%%%%%%
%%%%%%%%%%%%%%%%%%%%%%%%%%%%%%%%%%%%%%%%%%%%%%%%%%%%%%%%%%
\section{Group Manifolds in Matrix Model 
}\label{appendix:stability_instability}
%%%%%%%%%%%%%%%%%%%%%%%%%%%%%%%%%%%%%%%%%%%%%%%%%%%%%%%%%%
%%%%%%%%%%%%%%%%%%%%%%%%%%%%%%%%%%%%%%%%%%%%%%%%%%%%%%%%%%
%%%%%%%%%%%%%%%%%%%%%%%%%%%%%%%%%%%%%%%%%%%%%%%%%%%%%%%%%%
\hspace{4.6mm}
In this section, we discuss the covariant derivatives 
on group manifolds. 
We begin with the discussion for general group $G$ 
and then discuss the case of $SU(2)$ in detail. 
%%%%%%%%%%%%%%%%%%%%%%%%%%%%%%%%%%%%%%%%%%%%%%%%%%%%%%%
%%%%%%%%%%%%%%%%%%%%%%%%%%%%%%%%%%%%%%%%%%%%%%%%%%%%%%%
%%%%%%%%%%%%%%%%%%%%%%%%%%%%%%%%%%%%%%%%%%%%%%%%%%%%%%%
\subsection{Covariant Derivatives on Group Manifolds
}\label{subsec:general group manifold}
%%%%%%%%%%%%%%%%%%%%%%%%%%%%%%%%%%%%%%%%%%%%%%%%%%%%%%%
%%%%%%%%%%%%%%%%%%%%%%%%%%%%%%%%%%%%%%%%%%%%%%%%%%%%%%%
%%%%%%%%%%%%%%%%%%%%%%%%%%%%%%%%%%%%%%%%%%%%%%%%%%%%%%%
\hspace{4.6mm}
In this subsection, we consider the covariant derivative 
on a group manifold $G$. 
As shown in preceding sections, 
the left-invariant derivative can be identified with 
the covariant derivative $D_a$ with 
an appropriate field strength of $B$.  
The Hilbert space $V$ on which $D_a$ acts is 
the set of functions on the principal 
$Spin(d)$-bundle over $G$, where $d=\mbox{dim} G$. 
As we can see from (\ref{vielbein_2}),
the principal bundle can be covered by a single patch 
(i.e. the index $a$ in (\ref{vielbein_2}) is defined globally), 
and hence this bundle is the direct product 
of $G$ and $Spin(d)$. Therefore, we have 
\begin{eqnarray}
  V
  =
  C^\infty(G\times Spin(d))
  =
  C^\infty(G)\otimes
  C^\infty(Spin(d)). 
\end{eqnarray}
From (\ref{definition of left inv deriv}), 
$D_a$ acts on the first component 
$C^\infty(G)$ as the left-invariant derivative, or equivalently 
as the regular representation. 
If we denote the regular representation of the generator $t_a$ by 
$T_a^{(reg)}$, $D_a$ can be expressed as 
\begin{eqnarray}
  D_a=-iT_a^{(reg)}\otimes\textbf{1}. 
\end{eqnarray}
Using the standard procedure explained in 
section \ref{sec:Matrix representation of covariant derivative}, 
the index $a$ can be rewritten to that with a parenthesis:  
\begin{eqnarray}
  D_{(a)}
  =
  R_{(a)}{}^b(g^{-1})D_b
  =
  -iT_b^{(reg)}\otimes R_{(a)}{}^b(g^{-1}).
  \label{cov deriv on G with index with parenthesis} 
\end{eqnarray}

A remark is in order here. Although $D_a$ satisfies the commutation 
relation of the Lie algebra, $D_{(a)}$ does not for 
$G\neq SU(2)=Spin(3)$ 
because the structure constant $f_{abc}$ is not invariant under 
the action of $Spin(d)$. Indeed, the commutation relation of $D_{(a)}$ is 
\begin{eqnarray}
  [D_{(a)},D_{(b)}]
  &=&
  R_{(a)}{}^{a^\prime}(g^{-1})
  R_{(b)}{}^{b^\prime}(g^{-1})
  [D_{a^\prime},D_{b^\prime}]
  \nonumber\\
  &=&
  R_{(a)}{}^{a^\prime}(g^{-1})
  R_{(b)}{}^{b^\prime}(g^{-1})
  f_{a^\prime b^\prime c^\prime}
  D_{c^\prime}
  \nonumber\\
  &=&
  R_{(a)}{}^{a^\prime}(g^{-1})
  R_{(b)}{}^{b^\prime}(g^{-1})
  R_{(c)}{}^{c^\prime}(g^{-1})
  f_{a^\prime b^\prime c^\prime}
  D_{(c)}, 
\end{eqnarray}
where $g\in SU(2)$, 
and the coefficient 
$R_{(a)}{}^{a^\prime}(g^{-1})
R_{(b)}{}^{b^\prime}(g^{-1})
R_{(c)}{}^{c^\prime}(g^{-1})
f_{a^\prime b^\prime c^\prime}$ 
depends on $g$ unless $G=SU(2)$. 
In the case of $G=SU(2)$, this coefficient does not depend 
on $g$, 
\begin{eqnarray}
  R_{(a)}{}^{a^\prime}(g^{-1})
  R_{(b)}{}^{b^\prime}(g^{-1})
  R_{(c)}{}^{c^\prime}(g^{-1})
  \epsilon_{a^\prime b^\prime c^\prime}
  =
  \epsilon_{abc}, 
\end{eqnarray}
and $D_{(a)}$ also forms a Lie algebra.  
Therefore, we can find the explicit form of 
$D_{(a)}$ in terms of matrices.  
%%%%%%%%%%%%%%%%%%%%%%%%%%%%%%%%%%%%%%%%%%%%%%%%%%%%%%%
%%%%%%%%%%%%%%%%%%%%%%%%%%%%%%%%%%%%%%%%%%%%%%%%%%%%%%% 
%%%%%%%%%%%%%%%%%%%%%%%%%%%%%%%%%%%%%%%%%%%%%%%%%%%%%%%
\subsection{Explicit Form of the Covariant Derivative 
  on $SU(2)$ Group Manifold
}\label{appendix:S3_matrix}
%%%%%%%%%%%%%%%%%%%%%%%%%%%%%%%%%%%%%%%%%%%%%%%%%%%%%%% 
%%%%%%%%%%%%%%%%%%%%%%%%%%%%%%%%%%%%%%%%%%%%%%%%%%%%%%% 
%%%%%%%%%%%%%%%%%%%%%%%%%%%%%%%%%%%%%%%%%%%%%%%%%%%%%%%
\hspace{4.6mm}
In this subsection, we show the explicit form of the covariant derivative 
on group manifold $SU(2)$ in terms of matrices. 
Let $D_{a}^\prime$ be the covariant derivative with torsion 
which is obtained from $D_a$, 
\begin{eqnarray}
  D_{a}=\nabla_{a}+\frac{1}{2}\epsilon_{abc}{\cal O}_{bc},  
\end{eqnarray}
by flipping the sign of the torsion: 
\begin{eqnarray}
  D_{a}^\prime=\nabla_{a}-\frac{1}{2}\epsilon_{abc}{\cal O}_{bc}. 
\end{eqnarray}
The basic observation is that $D_{(a)}$ and $D_{(a)}^\prime$ 
commute with each other and that both of them 
satisfy the commutation relation of $SU(2)$:  
\begin{eqnarray}
  [D_{(a)}, D_{(b)}^\prime]=0, 
  \qquad
  [D_{(a)}, D_{(b)}]=\epsilon_{abc}D_{(c)}, 
  \qquad
  [D_{(a)}^\prime, D_{(b)}^\prime]
  =-\epsilon_{abc}D_{(c)}^\prime. 
\end{eqnarray}
Therefore, using an appropriate basis 
we can write $D_{(a)}$ and $D_{(a)}^\prime$ as 
\begin{eqnarray}
  & &
  D_{(a)}=-iJ_a^{(reg)}\otimes\textbf{1}, 
  \label{explicit form for SU(2)}
  \\
  & &
  D_{(a)}^\prime=\textbf{1}\otimes iJ_a^{(reg)},  
\end{eqnarray}
where $J^{(reg)}$ is the regular representation of $SU(2)$. 
The explicit form of $J^{(reg)}$ is given by 
\begin{eqnarray}
  J^{(reg)}
  =
  \left(
    \begin{array}{ccccc}
      J_{[0]} & & & &  \\
      & J_{[1/2]}\otimes\textbf{1}_2 & & &  \\
      & & \ddots & &  \\
      & & & J_{[l]}\otimes\textbf{1}_{2l+1} &  \\
      & & & & \ddots
    \end{array}
  \right),  
\end{eqnarray}
where $J_{[l]}$ stands for the spin-$l$ representation. 
Using (\ref{explicit form for SU(2)}), we can express 
the covariant derivative without torsion $\nabla_{(a)}$ 
and the Lorentz generator ${\cal O}_{ab}$ as 
\begin{eqnarray}
  & &
  \nabla_{(a)}
  =-
  \frac{i}{2}
  \left(
    J_a^{(reg)}\otimes\textbf{1} 
    -
    \textbf{1}\otimes J_a^{(reg)}
  \right), 
  \\
  & &
  {\cal O}_{ab}
  =-
  \frac{i}{2}
  \epsilon_{abc}
  \left(
    J_c^{(reg)}\otimes\textbf{1}
    +
    \textbf{1}\otimes J_c^{(reg)}
  \right). 
\end{eqnarray}

%%%%%%%%%%%%%%%%%%%%%%%%%%%%%%%%%%%%%%%%%%%%%%%%%%%%%%%%%%
%%%%%%%%%%%%%%%%%%%%%%%%%%%%%%%%%%%%%%%%%%%%%%%%%%%%%%%%%%
%%%%%%%%%%%%%%%%%%%%%%%%%%%%%%%%%%%%%%%%%%%%%%%%%%%%%%%%%%
%%%%%%%%%%%%%%%%%%%%%%%%%%%%%%%%%%%%%%%%%%%%%%%%%%%%%%%%%%
%%%%%%%%%%%%%%%%%%%%%%%%%%%%%%%%%%%%%%%%%%%%%%%%%%%%%%%%%%
%%%%%%%%%%%%%%%%%%%%%%%%%%%%%%%%%%%%%%%%%%%%%%%%%%%%%%%%%%
\vspace{0.4cm}

In the latter part of this subsection, 
we discuss whether or not the background 
$A_a=iD_{(a)}$ given in the previous subsection is stable  
in three-dimensional matrix models. 
Although the physical interpretation is different, 
this background is mathematically the same as 
multiple fuzzy sphere configuration arranged in a concentric pattern, 
whose dynamical properties have been studied extensively.   
In the model defined by the action (\ref{action_with_cubic_term}), 
this configuration has flat directions 
$J_{[l]}^a \to J_{[l]}^a+c^a\cdot \textbf{1}_{2l+1}$, 
where $c^a$ is a constant  
which represents the translation of each fuzzy sphere, 
and furthermore, 
if $c_a$ takes a non-zero value,
then tachyonic modes arise and 
the multiple fuzzy sphere configuration decays to a single fuzzy sphere  
which is described by the irreducible representation \cite{JMWY}. 
Therefore, the background considered here is unstable in this case. 
This instability may be removed by introducing supersymmetry 
or by adding a mass term.
For example, the three-dimensional supersymmetric matrix model 
with the cubic term is considered in \cite{IKTW} 
and it was shown that multiple fuzzy sphere configuration is BPS 
if they are arranged in a concentric pattern. 
Therefore, the configuration (\ref{explicit form for SU(2)}) 
would be protected from the quantum correction and be stable, 
while other configurations which are shifted with nonzero $c^a$ 
are not.  
In the case that the action has the mass term, 
it is easy to see that there is no flat direction. 
In \cite{ABN}, three-dimensional matrix model 
with the cubic and the positive mass term is considered 
at one-loop level, and it was shown that the multiple fuzzy sphere 
configuration becomes the true vacuum.\footnote{ 
In \cite{AKS}, this model was studied further by using 
the improved perturbation theory, and it was argued that 
this vacuum is indeed the nonperturbative one.  
} 
The multiplicity of each irreducible representation depends on 
the mass and the coefficient of the cubic term, and 
it is interesting to investigate whether 
the configuration (\ref{explicit form for SU(2)}) can be 
a true vacuum. 
%%%%%%%%%%%%%%%%%%%%%%%%%%%%%%%%%%%%%%%%%%%%%%%%%%%%%%%%%%
%%%%%%%%%%%%%%%%%%%%%%%%%%%%%%%%%%%%%%%%%%%%%%%%%%%%%%%%%%
%%%%%%%%%%%%%%%%%%%%%%%%%%%%%%%%%%%%%%%%%%%%%%%%%%%%%%%%%%

\end{document}